\documentclass{article}

\usepackage{amssymb}
\usepackage{amsmath}
\usepackage{latexsym}
\newtheorem{theorem}{Theorem}
\newtheorem{lemma}[theorem]{Lemma}
\newtheorem{proposition}[theorem]{Proposition}
\newtheorem{corollary}[theorem]{Corollary}
\newtheorem{fact}[theorem]{Fact}

\newenvironment{proof}[1]{%
   \begin{trivlist}{}{\setlength{\topsep}{0cm}\setlength{\partopsep}{0cm}}
   \item \textbf{#1.\@}\hspace*{1ex}\ignorespaces}%
   {\makebox[0cm]{}\nolinebreak\hfill$\Box$\end{trivlist}}

\def\T{{\dagger}}
\def\imagi{{\imath}}
\def\del{\partial}
\def\norm#1{{\left\|{#1}\right\|}}
\def\glb1q{\diamond}
\def\Orbits{{\cal O}}
\def\orb{{\cal C}}
\def\Dom{\Omega}

\def\Z{{\mathbb Z}}

\def\R{{\mathbb R}}
\def\C{{\mathbb C}}

\def\Uham{X}
\def\Vham{Y}

\begin{document}

\title{Quantum Computing with Global One-and Two-Qubit Gates}
\author{
{G\'abor Ivanyos
\thanks{
Computer and Automation Research Institute
Hungarian Academy of Sciences (MTA SZTAKI),
Kende u. 13-17, H-1111 Budapest, Hungary.
E-mail:{\tt Gabor.Ivanyos@sztaki.hu}
}
}
\and
{Attila B. Nagy
\thanks{
MTA SZTAKI and
Budapest University of Technology and Economics (BME),
M\H{u}egyetem rkp. 3-9, H-1111 Budapest, Hungary.
E-mail: {\tt nagy@math.bme.hu}
}
}
\and
{Lajos R\'onyai
\thanks{
MTA SZTAKI and BME.
E-mail: {\tt ronyai@sztaki.hu}
}
}
}

\maketitle

\begin{abstract}
We present generalized and improved constructions
for simulating quantum computers with a polynomial slowdown
on lattices composed of qubits on which certain global versions
of one- and two-qubit operations can be performed.
\end{abstract}

\section{Introduction}

In \cite{IMN05} we have shown that usual quantum
circuits can be efficiently simulated on several
regions of lattices composed of qubits on which
instead of individual addressing of qubits
we can perform certain global versions of
one and two-qubit gates. The Hamiltonian of the
global version of a two-qubit gate acting
on a pair of a qubit is the (weighted) average
of the translates of the Hamiltonian of
the two-qubit gate. In \cite{IMN05} translations 
by lattice vectors were considered. A brief
discussion and references to experimental
results regarding implementation of
global gates can also be found in \cite{IMN05}.
Our simulation requires a constant blowup
of space and the slowdown is polynomial.

Global two-qubit Hamiltonians considered
in \cite{IMN05} and in this paper are special
cases of Hamiltonians built from pairwise
interactions. Results regarding
efficient simulation of all such Hamiltonians
using a fixed one with connected interaction
graph and local one-qubit operations can
be found e.~g.~in \cite{DNBT} (or in 
\cite{LBCF} and \cite{LCYY} in the context of
NMR quantum computers). In \cite{KWC}, 
the sets of operations which are 
implementable as products of global 
gates from various collections are 
described. In the latter manuscript
efficiency is not addressed:
products of arbitrary length
are allowed.

Methods not requiring individual addressability of local spins 
are given in a series of papers by Benjamin and his collaborators. 
In \cite{BB03} it is shown how different levels of addressability 
assumptions can be dispensed with in the case of a chain of spins.
Multidimensional generalisations and issues related to physical implementations
including error correction techniques are discussed  in \cite{Ben,BBK,BB04}. 
In an intermediate proposal, given in \cite{BB03}, spins only at 
even and odd positions have to be distinguished, but there 
is also a method that -- like the proposals in \cite{IMN05}
and the present paper -- does not require that distinction either. Note
however, that the endpoints of the chain play an essential
role in the proposal given in \cite{BB03}. In this paper 
symmetry receives a more attentive treatment and
we do not make use of edge effects. The most closely
related work to the present paper is probably \cite{VC05}, 
where translation and reflection symmetric global operations in a chain
of 5-level local systems were considered.
 
In this paper we strengthen the results
of \cite{IMN05} in two directions.
First, we extend the method of \cite{IMN05}
to more general models of global gates.
In these models, the global Hamiltonians
are built from translates of local
ones by elements of groups more general
than translation groups. In case of multidimensional
lattices these groups may include not only translations
but also transformations like rotations and reflections.
Most notably, we propose constructions
for the case of hypercubic lattices and
global gates acting simultaneously on
pairs of qubits at various Euclidean
distances. In contrast to \cite{IMN05}, 
the constructions in the present paper
do not exploit that the domain containing
the lattice points we are working with has 
borders.

Another direction of improvement is 
restricting the set of global two-qubit
gates needed for the simulation. We show
that efficient simulation is
possible with global one-qubit gates and 
global versions of two-qubit gates whose 
Hamiltonians are diagonal in the computational
basis. As the translates of such gates
commute, in these cases the global gate
is just the simultaneous action of the
local translates. We note that some of the
global gates in \cite{IMN05} are not of
this form and therefore it seems to be much
harder to implement them in practice.

It is also natural to restrict the 
the "distances" of pairs of qubits
which we let global gates act on.
The most natural restriction would 
be allowing only next-neighbour interactions
(e.g,, distance $1$). Results with next-neighbor interactions
can be found in \cite{R05} and 
\cite{VC05}. The latter paper presents
a method which works with next-neighbor 
interactions in lattices composed of
5-level systems rather than qubits.
The method of \cite{R05} works
with a chain of qubits but it
exploits that the chain has endpoints.
It is not known if there is a version
of our method using only global
two-qubit gates with distance $1$.
In the one-dimensional case
we can present a version
in which the global two-qubit
gates have distance at most 22. It
works with qubits and does not
make use of the border of
the chain. Furthermore, our method
also works if there is certain
imperfection in the global gates.

The constructions of the present paper
require an initial state where qubits
at certain positions are set to one
while the others are set to zero. Obviously,
form the all-zero state such state cannot
be achieved using global gates. In \cite{VC05}
this problem is circumvented using certain
techniques such as running several reflection 
symmetric simulations at sufficiently large distances
simultaneously. Such techniques seem to be applicable 
in the case of 2-level (qubit) systems as well -- maybe
using a bigger system of global two-qubit gates
in the initialization phase.

The structure of this paper is the following. 
The model of global gates is described
in Section~\ref{globgates-sect}.
Our main theorem is stated
in Section~\ref{simu-sect},
where also combinatorial
notions describing the schemes
making efficient simulation
possible are introduced. 
Section~\ref{schemes-sect}
is devoted to examples
for such schemes,
while the proof of our main result
can be found in Section~\ref{extract-sect}.
We present a construction for
simulation using only global two-qubit
gates with distances between 1 and 22 in
Section~\ref{shift-sect}.

\section{Global gates}
\label{globgates-sect}

In this section we introduce some notation and 
describe a general mathematical model for global one- and 
two-qubit gates.

Throughout the paper we assume that
$G$ is a transitive permutation group
on a possibly infinite set $\Dom$
and $D$ is a subset of $\Dom$.
For $p\in \Dom$ and $g\in G$
we denote by $p^g$ the image of $p$ under $g$.
The permutation action of $G$ induces an equivalence 
relation $\sim$ on $D\times D$: $(p,q)\sim(p',q')$ 
if there is an element $g\in G$ such that $p'=p^g$ 
and $q'=q^g$.
We denote by $\Orbits$ the set of the
equivalence classes of $\sim$ different
from the diagonal $\{(p,p)|p\in D\}$.
For $(p,q)\in D\times D$ the equivalence class 
containing $(p,q)$ is denoted by $\orb_{(p,q)}$.

We encourage the reader to consider
the following instructive class of examples.
Here $\Dom=\R^s$, the $s$-dimensional Euclidean 
space and $G$ is the Euclidean group $E(s)$
which consists of isometries of $R^s$. For every
subset $D\subseteq \R^s$, two pairs $(p,q),(p',q')$
of $D\times D$ are equivalent under the action of $E(s)$
if and only if the distances $|p-q|$ and $|p'-q'|$
are the same. As a consequence, the classes of $\Orbits$
can be indexed by positive real numbers (distances) $\delta$ and
$(p,q)\in \orb_\delta$ if and only if $|p-q|=\delta$.
(Of course, depending on the choice of $D$, 
the class $\orb_\delta$ may be empty for certain $\delta$'s.
We shall refer to this geometric example in
intuitive explanations of abstract notions and arguments.

To define global gates,
we assume that $D$ is finite with
$|D|=n$ and we have a configuration
of $n$ qubits sitting at
each element of $D$. The Hilbert space 
of the pure states over these $n$ qubits
is $\C^{2^n}$. 
The elements of the standard basis
are indexed by the functions $a:D\rightarrow\{0,1\}$.
In order to shorten notation, for $p\in D$ we also
write $a_p$ for the value $a(p)\in\{0,1\}$.

As a part of our model we introduce a balance 
function $W:D\times D\rightarrow \R_{>0}$
on the pairs of $D$. This function will make it
possible to model effects like imperfection of global gates,
namely that the tool performing a global gate 
(e.g, a laser beam or a magnetic field) may act with
different strength on pairs or singletons of
qubits at different positions. Of course we
could take $W$ to be constant. This choice
would express invariance of global gates
under the action of $G$ perfectly. 
In order to shorten notation we will also denote $W(p,p)$
by $W(p)$.

Below we give a formal definition of
the model of global gates. The first
part is devoted to describe how global
Hamiltonians are built from local ones.
Then global gates are obtained from
global Hamiltonians in the usual way.
As in Section~\ref{extract-sect} we
shall make use of the formalism
introduced here in the context
of arbitrary operations, in the
first part we do not assume Hermiticity
of operations. That is, by an operation
we mean an arbitrary linear transformation
or matrix of the appropriate dimension. 
However, the reader not interested
in the details given in Section~\ref{extract-sect}
may think merely of Hamiltonians first.

For a one-qubit operation or
$2\times 2$ matrix $M$, whose rows
and columns are indexed by $0$ and $1$,
and an element $p\in D$ we write
$M^p$ for the $n$-qubit operation
which acts as $M$ on the qubit
at position $p$:
$$M^{p}_{a,b}=
\left\{\begin{array}{ll}
M_{a_p,b_p} & 
\mbox{if $a_s=b_s$ for every $s\in D\setminus\{p\}$,}
\\ 0 & \mbox{otherwise.} 
\end{array}\right.
$$
We will refer to $M^p$ as a {\it local one-qubit operation}
at position $p$. The {\it global one-qubit operation}
corresponding to $M$ is
$$M^\glb1q=\sum_{p\in D}W(p)M^{p}.$$
Similarly, for a 2-qubit operation or 
$2^2\times 2^2$ matrix $M$ (rows and columns
indexed by $00,01,10,11$) and a pair of elements 
$p\neq q\in D$, the {\it local two-qubit operation} $M^{(p,q)}$ 
at position $(p,q)$ is defined as
$$M^{(p,q)}_{a,b}=
\left\{\begin{array}{ll}
M_{(a_p,a_q),(b_p,b_q)} & 
\mbox{if $a_s=b_s$ for every $s\in D\setminus\{p,q\}$,}
\\ 0 & \mbox{otherwise,} 
\end{array}\right.
$$
and for a $\orb\in\Orbits$ the {\it global two-qubit operation}
is
$$M^\orb=
\sum_{(p,q)\in\orb}W(p,q)M^{(p,q)}.$$

A \textit{two-qubit global Hamiltonian} is a matrix of the
form $H^{\orb}$ where $H$ is an Hermitian 2-qubit
operation and a \textit{global 2-qubit gate} is an
operation of the form $\exp({-\imagi H^{\orb}})$ where
$H^{\orb}$ is a global two-qubit Hamiltonian. Global
one-qubit gates are of the form 
$\exp({-\imagi H^{\glb1q}})$ where $H$ is an
Hermitian one-qubit operation.

Note that global one-qubit gates can be interpreted
as parallel execution of local operations since
$$\exp({-\imagi H^{\glb1q}})=\prod_{p\in D}
\exp({-\imagi W(p)H^{p}}),$$
where the product on the right hand side can be taken
in an arbitrary order of the terms. The
analogous statement for global two-qubit operations
is not true in general. Only if for all $(p,q),(p',q')\in\orb$
the local operators $H^{(p,q)}$ and $H^{(p',q')}$
commute (this happens for example if $H$ is diagonal
in the standard basis) can  $\exp({-\imagi H^{\orb}})$
be decomposed as 
$\prod_{(p,q)\in \orb}\exp({-\imagi W(p,q)H^{(p,q)}}).$

\section{Simulation results}
\label{simu-sect}

An intuitive description
of our method for simulating a "usual"
quantum computation with global
gates acting on $D$ is the following.
We will designate two disjoint subsets
$P$ and $R$ of $D$. The qubits at positions
in $R$ will serve as reference points. Before
and after each logical step of the simulation,
their value will be set to one. Similarly,
the values of the qubits at positions {\it outside} 
$P$ and $R$ will be set to zero. The subset
$P$ will be the workspace. The qubits at positions
in $P$ can take arbitrary values and in each logical
step we simulate a local one- or two-qubit gate
at certain (pairs of) positions in $P$. To be more
precise, each logical step will {\it approximate}
such a gate.

By approximation we mean approximation in terms
of the {\it operator norm}. For an operator $U$ 
on the Hilbert space $\C^{2^n}$ we denote by 
$\norm{U}$ the operator
norm of $U$: $\norm{U}=\sup_{|x|=1}|Ux|$.
Note that $\norm{AB}\leq \norm{A}\cdot\norm{B}$.
Furthermore, if $\norm{A_1},\ldots,\norm{A_N},
\norm{B_1},\ldots,\norm{B_N}\leq 1$
(this holds in particular if $A_i$ and $B_i$
are unitary operators), then we have
\begin{equation}
\label{error-add}
\norm{A_1\cdots A_N-B_1\cdots B_N}
\leq \sum_{j=1}^N\norm{A_j-B_j}.
\end{equation}
We say that $A$ $\epsilon$-approximates $B$
if $\norm{A-B}\leq \epsilon$. Equation (\ref{error-add})
implies that in order to $\epsilon$-approximate
a circuit consisting of $\ell$ gates it is sufficient
to $\epsilon/\ell$-approximate each gate occurring in
the circuit.

Below is a very informal -- metaphoric -- description of our
simulation method. In this description qubits correspond to two-state
("up" or "down") switches placed at positions in $D$. The global
one-qubit gates are be modelled by the ability
of flipping the state of the switches simultaneously. 
The additional device corresponding to the two-qubit
gates with distance $\delta$ is the following.
For every $p\in D$, this tool {\it commits} the change of 
switch at $p$ if there is a position $q$ at distance
$\delta$ such that the original state of the switch
at $q$ was "up". The {\em value} of the commitment is
defined as the number of such points $q$. 
After a sequence of commitment
operations the change is performed at positions
where it has been committed by all of the commitments.

We work with states where the switches at points
in $R$ will be always "up", switches at points
in $P$ can be both "up" and "down", while the rest
are always "down". The goal is to be able switching
the state of a specific switch in $p\in P$ by keeping
the state of the others using the operations
described above. We also assume that this must
be done performing the global flip operation
followed by a sequence of commitments which
depend only on the position $p$ and not on the
particular state of the switches in $P$.
Furthermore, we require that for committed flips,
the total value of commitments does not depend
either on the particular state. To perform this 
we need switches at $r_1,\ldots,r_k$ in $R$ such that in any case,
commitment operations with distances $|p-r_1|,\ldots,|p-r_k|$ 
commit the flip only at position $p$. That is,
we need to be able to uniquely locate point 
$p\in P$ in such a strong sense just by the 
distances of it form points of $R$. We shall
use the term "addressable" to express this property.

We stress that the description above is only metaphoric, 
looking for any direct connection with the reality would
be misleading. The actual and accurate description
of the simulation method can be found in the rather
technical Section~\ref{extract-sect}. Below we give the 
formal definitions.

We call a pair $(P,R)$ of disjoint subsets of $D$
a {\it workspace--base scheme}. 
We say that the workspace--base scheme $(P,R)$ {\it admits}
the function $a:D\rightarrow\{0,1\}$ (or $a$ is {\it admissible}
if $(P,R)$ is clear from the context) if 
$a_r=1$ for every $r\in R$ and $a_q=0$ for every
$q\in D\setminus (P\cup R)$. We say
that $p\in P$ is {\it addressable} 
by $r_1,\ldots,r_k\in R$ if
for every $k+1$-tuple
$(p',r_1',\ldots,r_k')\in D\times(P\cup R)^k$,
the property that 
for each $i=1,\ldots,k$
  either $(p',r_i')\in \orb_{(p,r_i)}$ or $(r_i',p')\in \orb_{(p,r_i)}$
for each $i=1,\ldots,k$ implies 
$p'=p$ and $r_1'\ldots,r_k'\in R$.
We say that $p\in P$ is $k$-addressable
if there exist $r_1,\ldots,r_k\in R$
such that $p$ is addressable by
$r_1,\ldots,r_k$. We call $(P,R)$ $k$-addressable if
every $p\in P$ is $k$-addressable.

The subspace of $\C^{2^n}$ spanned by the basis elements
corresponding to the admissible functions is
obviously isomorphic to $\C^{2^{|P|}}$. Its elements
(of length 1) are called admissible vectors
(or states, respectively). The 
remaining basis elements will span the orthogonal
complement of the subspace of admissible states.
The vectors (or states) in the latter subspace 
will be called inadmissible. Our first (and
main) goal is to
produce unitary operations
which (approximately) preserve the subspace
of admissible vectors and, restricted to
this subspace, (approximately) act as local
one-qubit gates. In other words, if we adopt an order of 
the basis where the first $2^{|P|}$ basis elements 
correspond to the admissible functions and the 
rest correspond to the inadmissible functions,
we will produce operations whose matrices
are (approximately) block diagonal and 
have upper left $2^{|P|}\times 2^{|P|}$
block which is (approximately) the same
as the matrix of a local one-qubit
gate acting on a qubit at position $p\in P$.
Then, having the local one-qubit gates at hand
in order to get a universal quantum computer
on on the admissible states,
it is sufficient to use a unitary operator
which has an all-pair entangling Hamiltonian
on the admissible states. 

We shall achieve our main goal by taking
commutators of certain global one-qubit 
Hamiltonians with Hamiltonians of the form $A^\orb$
at the Lie algebra level,
where $\orb\in\Orbits$ and $A$ is the Hamiltonian
of the so-called {\it controlled phase shift}
operation. The entries of the matrix of  
this two-qubit Hamiltonian are all zero
except in the lower right corner (corresponding to
the basis element indexed by $11$):
\begin{eqnarray}
\label{CPS-eq}
A=\left(\begin{array}{cccc}
0 & 0 & 0 & 0 \\
0 & 0 & 0 & 0 \\
0 & 0 & 0 & 0 \\
0 & 0 & 0 & 1 
\end{array}\right).
\end{eqnarray}
The corresponding global two-qubit {\it gates} are
\begin{eqnarray}
\label{addr-global-eq}
\exp\left(\left(\begin{array}{cccc}
0 & 0 & 0 & 0 \\
0 & 0 & 0 & 0 \\
0 & 0 & 0 & 0 \\
0 & 0 & 0 & -T\imagi
\end{array}\right)^{\orb_{(p,r)}}\right),\,\,\,(T\in \R,p\in P,r\in R),
\end{eqnarray}
where $\imagi$ stands for the imaginary unit $\sqrt{-1}$.
\\
We shall approximate the local one-qubit gates
by products of global two-qubit gates of the form (\ref{addr-global-eq})
and global one-qubit gates of the form
\begin{eqnarray}
\label{oneq-global-eq}
\exp\left(\left(\begin{array}{cc}
0 & z  \\
-{\overline z} & 0 
\end{array}\right)^{\glb1q}\right),\,\,\,(z\in \C).
\end{eqnarray}

In Section~\ref{extract-sect} we shall prove the following.

\begin{theorem}
\label{simu-thm}
Assume that for every $p',p'',q',q''\in D$ we
have $\frac{W(p',q')}{W(p'',q'')}\leq w$. Assume further
that $(P,R)$ is a workspace--base scheme on $D$.
Then, for every $k$-addressable $p\in P$,
and for every $0<\epsilon<1$, every operation which acts
on admissible states as a one-qubit gate at $p$
with Hamiltonian of norm bounded by a constant 
can be $\epsilon$-approximated 
by a product of $(nw/\epsilon)^{O(k^2)}$
global gates of type
(\ref{addr-global-eq}) and (\ref{oneq-global-eq}).
\end{theorem}

If the "distance" between two points $p,q\in P$
cannot occur as a "distance" between a point $p'$
in $P$ and another point $r$ in $R$, that is
\begin{equation}
\label{norep-eq}
\left(\orb_{p,q}\cup\orb_{q,p}\right)\cap \orb_{p',r}=\emptyset
\mbox{~~for every $p,q,p'\in P$ and $r\in R$,}
\end{equation}
then the global gate 
\begin{eqnarray}
\label{twoq-global2-eq}
\exp\left(\left(\begin{array}{cccc}
0 & 0 & 0 & 0 \\
0 & 0 & 0 & 0 \\
0 & 0 & 0 & 0 \\
0 & 0 & 0 & -T\imagi
\end{array}\right)^{\orb_{(p,q)}}\right)
\end{eqnarray}
acts on the admissible states
as a unitary operator whose
Hamiltonian entangles the qubits
at $p$ and $q$ (and also the pairs
of qubits in $P$ with the same "distance").
Global gates of this type, together
with the simulated one-qubit gates on 
admissible states, using the selective 
decoupling techniques
(see e.g. \cite{LCYY,LBCF,DNBT}), give an
efficient simulation of any quantum circuit.
We obtain the following.
\footnote{We are indebted to an anonymous
referee for suggesting the argument
above. This has made it possible to replace 
a much more complicated
argument -- and a somewhat weaker statement --
in a  preliminary version of this paper.}

\begin{corollary}
\label{main-coro}
Assume that for every for every $p',p'',q',q''\in D$ we
have $\frac{W(p',q')}{W(p'',q'')}\leq w$. Assume further
that $(P,R)$ is a $k$-addressable workspace--base scheme on $D$
with additional property (\ref{norep-eq}).
Then, for every $0<\epsilon<1$, the action of 
any quantum circuit of length $\ell$ on $P$ 
on admissible states can be $\epsilon$-approximated 
by a product of $(nw\ell/\epsilon)^{O(k^2)}$ global gates of type
$(\ref{addr-global-eq})$, $(\ref{oneq-global-eq})$
and $(\ref{twoq-global2-eq})$. 
\end{corollary}

Here by a quantum circuit we mean
a circuit built from 2-qubit gates,
that is a sequence of unitary 2-qubit
operations. As 2-qubit gates are universal 
(see e.g. \cite{DiVi}), any circuit in the 
sense of \cite{Deutsch}
can be efficiently simulated as well.

\section{Extracting local gates by taking commutators}
\label{extract-sect}

Our main tool will be taking commutators with
global two-qubit Hamiltionians of the form $A^\orb$,
where the matrix $A$ is as in (\ref{CPS-eq}).
This will be the tool analogous to the
commitment operation in the metaphoric description.
See Proposition~\ref{extract1-prop} for the effect
of an appropriate sequence of such commutations.
It is easy to see that, for every $\orb\in \Orbits$,
$A^\orb$ is a diagonal matrix whose entry at
position $a,a$ is
$$A^\orb_{a,a}=
\sum_{\mbox{
\tiny$\begin{array}{cc}(p,q)\in\orb\\ a_p=a_q=1\end{array}$}}
W(p,q).$$
It follows that for arbitrary $a,b:D\rightarrow\{0,1\}$,
\begin{eqnarray*}
[A^\orb,E_{a,b}] & = &
\left(\sum_{\mbox{
\tiny$\begin{array}{cc}(p,q)\in\orb\\ a_p=a_q=1\end{array}$}}
W(p,q)-
\sum_{\mbox{
\tiny$\begin{array}{cc}(p,q)\in\orb\\ b_p=b_q=1\end{array}$}}
W(p,q)\right)E_{a,b}
\end{eqnarray*}
where $E_{a,b}$ is the $2^{|D|}\times 2^{|D|}$ elementary matrix having one at position 
$a,b$ and zero elsewhere. For $a:D\rightarrow\{0,1\}$ we denote
by $\del_p{a}$ the function which takes value zero at $p$ and
coincides with $a$ on $D\setminus \{p\}$. That is, $\del_p{a}$
is obtained from $a$ by deleting the $p^{\mbox{th}}$ bit. 
Commutators of $A^\orb$ with elementary matrices of the 
form $E_{a,\del_p a}$ turn out to be of particular interest.
We have $[A^\orb,E_{a,\del_p a}]=0$ if $a_p=0$ and 
\begin{eqnarray}
\label{comm1-eq}
[A^\orb,E_{a,\del_p a}] & = &
\left(
\sum_{\mbox{
\tiny$\begin{array}{cc} q\\ (p,q)\in\orb\\ a_q=1\end{array}$}}
W(p,q)+
\sum_{\mbox{
\tiny$\begin{array}{cc} q \\ (q,p) \in\orb\\ a_q=1\end{array}$}}
W(q,p)\right)E_{a,\del_p a}
\end{eqnarray}
if $a_p=1$.

For $\orb\in\Orbits$ we define a modified balance
function $W_\orb$ on $D\times D\setminus\{(p,p)|p\in D\}$
by
$$W_\orb(p,q)=\left\{\begin{array}{ll}
0 & \mbox{if $(p,q)\not\in \orb$ and $(q,p)\not\in\orb$,}\\
W(p,q) & \mbox{if $(p,q)\in \orb$ and $(q,p)\not\in\orb$,} \\
W(q,p) & \mbox{if $(q,p)\in \orb$ and $(p,q)\not\in\orb$,} \\
W(p,q)+W(q,p) & \mbox{if $(p,q)\in \orb$ and $(q,p)\in\orb$.}
\end{array}\right.
$$
Iterated application of (\ref{comm1-eq}) 
gives the following two lemmas.

\begin{lemma}
\label{null-lemma}
Let $p\in D$, $\orb_1,\ldots,\orb_k\in\Orbits$,
and $a:D\rightarrow\{0,1\}$ with $a_p=1$. 
Assume further that there is
no $k$-tuple $q_1',\ldots,q_k'\in D$
such that $a_{q_1'}=\ldots=a_{q_k'}=1$ and
either $(p,q_i')\in \orb_i$
or $(q_i',p)\in \orb_i$ holds
for every $i=1,\ldots,k$. Then
\begin{eqnarray*}
[A^{\orb_1},[\ldots,[A^{\orb_k},E_{a,\del_p a}]]]=0
\end{eqnarray*}
\end{lemma}

\begin{lemma}
\label{ext0-lemma}
Let $p\in D$, $\orb_1,\ldots,\orb_k\in\Orbits$,
and $a:D\rightarrow\{0,1\}$ with $a_p=1$. Let
$Q\subseteq D\setminus\{p\}$ such that
$a_q=1$ for every $q\in Q$ and
$q_1,\ldots,q_k\in Q$ such that
$(p,q_i)\in \orb_i$ or $(q_i,p)\in\orb_i$
for $i=1,\ldots,k$. 
Assume further that
if for a $k$-tuple $q_1',\ldots,q_k'\in D$
we have $a_{q_1'}=\ldots=a_{q_k'}=1$ and
either $(p,q_i')\in \orb_i$
or $(q_i',p)\in \orb_i$ holds
for every $i=1,\ldots,k$, then
$q_1',\ldots,q_k'\in Q$.
Then 
\begin{eqnarray*}
[A^{\orb_1},[\ldots,[A^{\orb_k},E_{a,\del_p a}]]]=
W_1\cdots W_k E_{a,\del_p a},
\end{eqnarray*}
where 
$W_i=\sum_{q\in Q}W_{\orb_i}(p,q)$.
\end{lemma}

Lemmas~\ref{ext0-lemma} and~\ref{null-lemma} imply
the following.

\begin{lemma}
\label{extract-lemma}
Let $(P,R)$ be the workspace--base scheme,
$a:D\rightarrow \{0,1\}$, $p\in P$ and $q\in D$.
Assume that $p$ is addressable by $r_1,\ldots,r_k\in R$.
Under these assumptions, if $a$ is admissible then
\begin{eqnarray*}
[A^{\orb_1},[\ldots,[A^{\orb_k},E_{a,\del_q a}]]]=
\left\{\begin{array}{ll}
W_1\cdots W_kE_{a,\del_q a}
  & \mbox{if $q=p$ and $a_p=1$,}
\\ 0 & \mbox{otherwise,}
\end{array}\right.
\end{eqnarray*}
where $\orb_i=\orb_{(p,r_i)}$ 
and $W_i=\sum_{r\in R}W_{\orb_i}(p,r)$
for $i=1,\ldots,k$. 
If $a$ is not admissible but $\del_q a$ is admissible
then
\begin{eqnarray*}
[A^{\orb_1},[\ldots,[A^{\orb_k},E_{a,\del_q a}]]]=0.
\end{eqnarray*}
\end{lemma}

\begin{proof}{Proof}
Assume that either $a$ or $\del_q a$ is admissible
and that the commutator is nonzero. Then $a_q=1$, and, 
by Lemma~\ref{null-lemma}, there exist
$r_1'\ldots,r_k'\in D$ such that 
$a_{r_1'}=\ldots=a_{r_k'}=1$ and
either $(q,r_i')\in\orb_i$ or $(r_i',q)\in\orb_i$.
Since $a$ or $\del_q a$ is admissible,
$r_i'\in P\cup R$ and from the definition
of workspace--base schemes it follows
that $q=p$ and $r_i'=r_i$ for $i=1,\ldots,k$.
This shows the second statement and part of the first one.
The rest follows from Lemma~\ref{ext0-lemma}.
\end{proof}

For an elementary one-qubit global
operation we have the following.

\begin{lemma}
\label{extract1-lemma}
Let $(P,R)$ be a workspace--base scheme,
$a,b:D\rightarrow \{0,1\}$, and assume that $p\in P$
is addressable by $r_1,\ldots,r_k$. Suppose further that
either $a$ or $b$ is admissible. 
Then, for the $2 \times 2$ matrix
$$B=\left(\begin{array}{cc} 0 & 1 \\ 0 & 0 \end{array}\right)
$$
we have 
\begin{eqnarray*}
[A^{\orb_1},[\ldots,[A^{\orb_k},B^\glb1q_{a,b}]]]=
W(p)W_1\cdots W_kB^p_{a,b},
\end{eqnarray*}
where $\orb_i=\orb_{(p,r_i)}$ 
and $W_i=\sum_{r\in R}W_{\orb_i}(p,r)$
for $i=1,\ldots,k$. 
\end{lemma}

\begin{proof}{Proof}
We denote by $C$ the commutator on the left hand side
of the asserted equality.
For every $q\in D$ we have
$B^{q}=
\sum_{a\mid a_q=1}
E_{a,\del_q a}
$. 
Therefore 
\begin{eqnarray*}
B^\glb1q & = &
\sum_{q\in D}W(q)
\sum_{a\mid a_q=1}E_{a,\del_q a}
=
\sum_{a}\sum_{q\mid a_q=1}W(q)E_{a,\del_q a}
\end{eqnarray*}
From this equality, using Lemma~\ref{extract-lemma},
we infer 
\begin{eqnarray*}
C_{a,b}=
\left\{\begin{array}{ll}
W(p)\prod_{i=1}^kW(p,r_i) & \mbox{if $b=\del_p(a)$ and $a_p=1$,} 
\\
0 & \mbox{otherwise,}
\end{array}\right.
\end{eqnarray*}
whenever either $a$ or $b$
is admissible. {}From the latter equality the assertion follows 
because $B^p_{a,b}$ is one if $a_p=1$, $b=\del_p a$;
and zero otherwise.
\end{proof}


For a global one-qubit Hamiltonian we obtain
the following.

\begin{proposition}
\label{extract1-prop}
Assume that $(P,R)$ 
is a workspace--base scheme on $D$,
and let 
$$\Uham=\left(
\begin{array}{cc} 0 & z \\ -{\overline z} & 0\end{array}\right),$$
where $z$ is a complex number. 
Suppose further that $p\in P$ is addressable by $r_1,\ldots,r_k\in R$. 
Then for every pair $a,b:D\rightarrow \{0,1\}$ such that
either $a$ or $b$ is admissible, we have
\begin{eqnarray*}
[-\imagi A^{\orb_1},[\ldots,[-\imagi A^{\orb_k},
\Uham^\glb1q]]]_{a,b}=
W(p)W_1\cdots W_k {\widetilde{\Uham}}^p_{a,b}, 
\end{eqnarray*}
where $\orb_i=\orb_{(p,r_i)}$,
$W_i=\sum_{r\in R}W_{\orb_i}(p,r)$,
and
$${\widetilde \Uham}=\left(
\begin{array}{cc} 0 & (-\imagi)^kz \\  -\imagi^k{\overline z} & 0 \end{array}\right).$$
\end{proposition}

\begin{proof}{Proof}
Observe that $\Uham=zB-\overline zB^\T$ where $B$ is the same
$2\times 2$ matrix as in Lemma~\ref{extract1-lemma}. 
Hence, using also that the matrices $A^{\orb_i}$
are self-adjoint, we obtain
\begin{eqnarray*}
[-\imagi A^{\orb_1},[\ldots,[-\imagi A^{\orb_k},
\Uham^\glb1q]]]
  = 
(-\imagi)^k[A^{\orb_1},[\ldots,[A^{\orb_k},
\Uham^\glb1q]]]
= \\
z(-\imagi)^k[A^{\orb_1},[\ldots,[A^{\orb_k},
B^\glb1q]]]
-{\overline z}(-\imagi)^k
[A^{\orb_1},[\ldots,[A^{\orb_k},
B^{\glb1q\T}]]]
= \\
z(-\imagi)^k[A^{\orb_1},[\ldots,[A^{\orb_k},
B^\glb1q]]]
-{\overline z}(-\imagi)^k
[A^{\orb_1},[\ldots,[A^{\orb_k},
B^\glb1q]]]^\T
\end{eqnarray*}
Using Lemma~\ref{extract1-lemma}, this equality gives
\begin{eqnarray*}
[-\imagi A^{\orb_1},[\ldots,[-\imagi A^{\orb_k},
\Uham^\glb1q]]]_{a,b} &=&
z(-\imagi)^k W(p)W_1\cdots W_k B^p_{a,b}
\\
& & 
-{\overline z}
(-\imagi)^k W(p)W_1\cdots W_k B^{\T p}_{a,b}
\\
& = &
W(p)W_1\cdots W_k {\widetilde{\Uham}}^p_{a,b}, 
\end{eqnarray*}
whenever either $a$ or $b$ is admissible.
\end{proof}

Proposition~\ref{extract1-prop} is
used to prove Theorem~\ref{simu-thm}, i.e.,
to show that the local one-qubit {\it gates}
can be efficiently approximated. We need the following 
fact on approximation of an operator
whose Hamiltonian is a commutator. For the proof,
see e.g., \cite{IMN05}.

\begin{fact}
\label{comm-approx-fact}
There is an absolute constant $c>0$, such that
\begin{eqnarray*}
\mbox{\huge$\|$}
\left(
\exp(-\frac{\imagi}{\mbox{\scriptsize$\sqrt{N}$}}\Uham^{-1})\cdot
\exp(-\frac{\imagi}{\mbox{\scriptsize$\sqrt{N}$}}\Vham^{-1})\cdot
\exp(-\frac{\imagi}{\mbox{\scriptsize$\sqrt{N}$}}\Uham)
\cdot
\exp(-\frac{\imagi}{\mbox{\scriptsize$\sqrt{N}$}}\Vham)
\right)^{N}
\\
-
\exp([-\imagi \Uham,-\imagi \Vham])
\mbox{\huge$\|$}
<
c\cdot M^3N^{-\frac{1}{2}}
\end{eqnarray*}
for any $N>M^2$, where
$\Uham$ and $\Vham$ are Hermitian operators on the
Hilbert space $\C^{2^n}$ and $M=\max\{\norm{\Uham},\norm{\Vham},1\}$. 
\end{fact}

We remark that commutators (both group and Lie theoretic,
in view of Fact~\ref{comm-approx-fact}) also play an 
important role in proof the
 Kitaev--Solovay theorem (see e.g., \cite{KSV}). It turns out that
 decomposing unitaries as group theoretic commutators can be used
to exponentially improving a sufficiently good
approximation of a unitary operator by a circuit built
from a given gate set.

Now we are in a position to prove Theorem~\ref{simu-thm}.

\begin{proof}{Proof of Theorem~\ref{simu-thm}}
We show the statement for gates of type
\begin{eqnarray}
\label{oneq-local-eq}
\exp\left(\left(\begin{array}{cc}
0 & z \\
-{\overline z} & 0 
\end{array}\right)^{p}\right),\,\,\,(z\in \C,|z|\leq 1).
\end{eqnarray}
Then, the general case for a general one-qubit gate
follows from Fact~\ref{comm-approx-fact} and Trotter's
formula since the third basis element
$$
\imagi\sigma_z=\left(\begin{array}{cc}
\imagi & 0 \\
0 & -\imagi  
\end{array}\right)
$$
of the Lie algebra $su_2$
equals $\frac{1}{2}$ of the commutator of 
the first two ones which are
$$\imagi\sigma_y=\left(\begin{array}{cc}
0 & 1 \\
-1 & 0 
\end{array}\right)
\mbox{~and~}
\imagi\sigma_x=\left(\begin{array}{cc}
0 & \imagi \\
\imagi & 0 
\end{array}\right).$$
Let 
$$Z=\frac{\imagi^{k+1}}{W(p)}
\left(
\begin{array}{cc}
0 & z \\
-{\overline z} & 0
\end{array}\right)^{\glb1q},
$$ and for $i=1,\ldots,k$ let 
$\Uham_i=1/W_i A^\orb_{(p,r_i)}$ where
$W_i=\sum_{r\in R}W_{\orb_{(p,r_i)}}(p,r)$. 
Then, by Proposition~\ref{extract1-prop},
on admissible states the operator
$\exp([\imagi \Uham_1,[\ldots[-\imagi \Uham_k,-\imagi Z]]])$
acts in the same way as the required one-qubit
local gate of type (\ref{oneq-local-eq}). For
$i=1,\ldots,k$ we denote the commutator
$[\imagi \Uham_i,[\ldots[-\imagi \Uham_k,-\imagi Z]]])$
by $\Vham_i$. Also set $\Vham_{k+1}=-\imagi Z$. Then
for $i=1,\ldots,k$ we have 
$\Vham_i=[-\imagi \Uham_i,\Vham_{i+1}]$. 
We have $\norm{Z}=O(nw)$, $\norm{\Uham_i}=O(nw)$
and $\norm{\Vham_i}=O((2nw)^{k-i+1})$.
It follows that all the norms
$\norm{\Vham_i}$ ($i=2,\ldots,k+1$)
$\norm{\Uham_i}$ ($i=1,\ldots,k$)
are bounded by $M=O(2nw)^k$.
Set $\epsilon_1=\epsilon$.
By Fact~\ref{comm-approx-fact},
we obtain an $\epsilon_i/2$-approximation
of $\exp{\Vham_i}$ by a product containing
$2N_i$ terms of $exp(\pm \Uham_i)$ and
$2N_i$ terms of $exp(\pm \Vham_{i+1})$,
where $N_i=O(M^6)/\epsilon_i$. If we
substitute each factor $\exp(\pm V_{i+1})$ by
approximations with error
$\epsilon_{i+1}=\epsilon_i/4N_i$,
the total error of the product will
be bounded by $\epsilon_i$. For
the sequence $\epsilon_i$ we obtain
a recursion $\epsilon_{i+1}=O(M^6\epsilon_i)$
($i=1,\ldots,k-1$). That is,
$\epsilon_i=(c'M^6)^{1-i}\epsilon$
for some constant $c'$ and hence
$N_i=O((c''M^6)^{i})$.
The total number of terms is
$4N_1+4N_1\cdot 4N_2+\ldots+4N_1\cdots 4N_k=
O(k(4c''M^6)^{k(k+1)/2})=(nw)^{O(k^2)}.$
\end{proof}

\section{Addressable workspace--base schemes}
\label{schemes-sect}

In this section we give some examples
of addressable workspace--base schemes
corresponding to various group actions.
In order to obtain nice ("uniform") constructions,
we extend the notion of addressability
(and connectedness) to possibly infinite
subsets $D\subseteq \Omega$ in the obvious way.

Assume that $(P,R)$ is a workspace--base scheme
in $D$ with $R=\{r_1,\ldots,r_k\}$. 
We say that $(P,R)$ 
is {\it strictly addressable} 
if for every
$p\in P$ and for every
$k+1$-tuple $(p',r_1',\ldots,r_k')\in D\times (P\cup R)^k$
$(p',r_i')\in \orb_{(p,r_i)}\cup\orb_{(r_i,p)}$ ($i=1,\ldots,k$)
implies $p'=p$ and $r_i'=r_i$ ($i=1,\ldots,k$).
This means that every $p\in P$ is addressable
by the whole of $R$ in a strict sense where -- in terms of our geometric
example -- the sequence of "distances" of $p$ from
$r_i$ identifies $p$ and $r_1,\ldots,r_k$.
We note that strict addressability implies condition
(\ref{norep-eq}) and therefore Corollary~\ref{main-coro}
is applicable to strictly addressable schemes.

\subsection{Schemes for translation-invariant global gates}

This part is devoted to constructions of strictly
addressable workspace--base schemes for translation-invariant
global gates in abelian groups, especially in the
$s$-dimensional lattice $\Z^s$. In particular, we shall show that
$(P_1,R_1)$ with $R_1=\{1,3,9\}$ and 
$P_1=4\Z=\{\ldots,-12,-8,-4,0,4,8,12,\ldots\}$
is a strictly addressable scheme for translations 
in $\Z$ of density $1/4$. 

The model of translations in $\Z$ can be interpreted
as that the qubits sit in a doubly infinite chain and 
the group acting on the chain is given by the shifts. 
Assuming a constant balance function, the global gates 
are shift-invariant. Just like in 
Section~\ref{globgates-sect}, the actual (finite) version
of the model is obtained by restricting the action
of the global operation to a finite section of the chain. 

For all positive integers $s$ the scheme above lifts to strictly 
addressable schemes $(P_s,R_s)$ where
$R_s=\{(1,0,\ldots,0),(3,0,\ldots,0),(9,\ldots,0)\}$
and $P_s=\{(4z_1,z_2,\ldots,z_s)|z_i\in \Z\}$. 
It is an open question whether $1/4$ is indeed the optimal
density or there are more efficient constructions.
We start with the formal definitions and give
some lemmas which may be useful in constructing
other addressable schemes.

Let $G$ be an abelian group. We use the additive notation
for the group operations. The group $G$ acts on $\Dom=G$
by shifting: $p^g=p+g$. We set $D=G$. As $(p,q)\sim (p',q')$ iff
$p'-q'=p-q$, the elements of $\Orbits$ can
be indexed by the elements of $G\setminus \{0\}$:
$(p,q)\in \orb_g$ iff $p-q=g$. In other words, 
$\orb_{(p,q)}=\orb_{p-q}$.

Under some mild restrictions, we can lift 
strictly addressable schemes from
factor groups as follows. 

\begin{lemma}
\label{lifting-lemma}
Assume that $\phi$ is a surjective homomorphism
from the abelian group $G$ onto $K$, and
$(Q,S=\{s_1,\ldots,s_k\})$ is a strictly addressable scheme
on $K$. Let $P=\phi^{-1}Q$. For each $i=1,\ldots,k$
we pick a single element $r_i$ from $\phi^{-1}\{s_i\}$.
Then, assuming that there exist $i,j\in \{1,\ldots k\}$
such that $2r_i\neq 2r_j$, the scheme
$(P,R=\{r_1,\ldots,r_k\})$ is a strictly addressable 
on $G$.
\end{lemma}

\begin{proof}{Proof}
Assume that $p'-r_i'=\pm(p-r_i)$ with $p\in P$, $p'\in G$,
$r_i'\in P\cup R$ ($i=1,\ldots,k$). Then 
$\phi(p')-\phi(r_i')=\pm(\phi(p)-\phi(r_i))$
and $\phi(p)\in Q$ and $\phi(r_i')\in Q\cup S$,
and strict addressability of $(Q,S)$ implies
$\phi(p')=\phi(p)$ and $\phi(r_i')=\phi(r_i)$ for $i=1,\ldots,k$.
From the latter facts and disjointness of $Q$ and $S$
we infer that $r_i'=r_i$ for $i=1,\ldots,k$.
Hence $p'-r_i=\pm (p-r_i)$ for $i=1,\ldots,k$.
If $p\neq p'$ then this is only possible
if $2r_i=p+p'$ for every $i$. 
\end{proof}

Lemma~\ref{lifting-lemma} gives several
straightforward ways of constructing strictly 
addressable schemes in multidimensional lattices
from one-dimensional schemes. In particular,
strict addressability of the scheme $(P_1,R_1)$
given above implies that of $(P_s,R_s)$ for
every $s$. 

As another application of Lemma~\ref{lifting-lemma}
we can give a simple construction for
a strictly addressable scheme of density $1/14$
in $\Z$.

\begin{proposition}
In $\Z_{14}$, the additive group of integers
modulo 14, the scheme $\{0\},\{1,3,7\}$ is a strictly
addressable workspace--base scheme. As
a consequence, so is the scheme $\{14z|z\in \Z\},\{1,3,7\}$
in $\Z$.
\end{proposition}

\begin{proof}{Proof}
Assume that for $p'\in \Z_{14}$ and
for $r_1',r_2',r_3'\in\{0,1,3,7\}$ we have
$p'-r_1'=\pm 1, p'-r_2'=\pm 3, p'-r_3'=\pm 7$.
If $p'$ is odd then $r_1',r_2',r_3'$ must be even
therefore $r_1'=r_2'=r_3'=0$ and
hence $\{\pm(p'-r_i')|i=1,2,3\}=\{\pm p\}$,
a set containing at most 2<3 elements,
a contradiction. Therefore $p'$ is even
and $r_1',r_2',r_3'$ must be odd which is
only possible if $\{r_1',r_2',r_3'\}\subseteq \{1,3,7\}$. 
On the other hand, 
$\{\pm(2-1),\pm(2-3),\pm(2-7)\}=\{1,13,9,5\}$,
$\{\pm(4-1),\pm(4-3),\pm(4-7)\}=\{3,11,1,13\}$,
$\{\pm(6-1),\pm(6-3),\pm(6-7)\}=\{5,9,3,11,13,1\}$,
$\{\pm(8-1),\pm(8-3),\pm(8-7)\}=\{7,5,9,1,13\}$,
$\{\pm(10-1),\pm(10-3),\pm(10-7)\}=\{9,5,7,3\}$
and none of these sets contain $\{1,3,7\}$, 
therefore $p'=0$. As $-1$ and
$-3$ are not in $\{1,3,7\}$, the only
possibility is that $r_1=1$, $r_2=3$ and $r_3=7$.
The second statement follows from Lemma~\ref{lifting-lemma}.
\end{proof}

We proceed with the somewhat tedious proof of
strict addressability of the scheme $(P_1,R_1)$
defined above. In order to support similar
but potentially more efficient constructions,
we give the proof in form of some statements.

Note that strictly addressable schemes
$(P,R)$  
where $P$ is non-empty and $|R|\leq 2$
do not exist in any abelian group $G$.
In the remainder of this subsection we 
consider schemes with
three-element bases.

\begin{lemma}
\label{refs-lemma}
Let $G$ be an abelian group,
$r_1,r_2,r_3\in G$ such that 
$r_i-r_j\neq r_{i'}-r_{j'}$ for
$(i,j)\neq (i',j')$. 
Then for every 
$p\in G$ such that $2p\not\in
\{{r_1+r_2},{r_1+r_3},{r_2+r_3}\}$ 
for every $p'\in G$,
for every $r_1',r_2',r_3'\in \{r_1,r_2,r_3\}$
the condition $p-r_i=\pm(p'-r_i')$
($i=1,2,3$) implies $p'=p$,
$r_1'=r_1$, $r_2'=r_2$ and $r_3'=r_3$.
\end{lemma}

\begin{proof}{Proof}
As $2p\not\in \{{r_1+r_2},{r_1+r_3},{r_2+r_3}\}$,
we have $p-r_i\neq \pm(p-r_j)$ for $i\neq j$.
Therefore $\{r_1',r_2',r_3'\}=\{r_1,r_2,r_3\}$
and there exists a permutation $\pi$ of $\{1,2,3\}$
and $\nu_1,\nu_2,\nu_3\in\{\pm 1\}$ such
that $p-r_i=\nu_i(p'-r_{i^\pi})$.
There exist $i\neq j$ such that $\nu_i=\nu_j=\nu$.
We have $p-r_i=\nu(p'-r_{i^\pi})$
and $p-r_j=\nu(p'-r_{j^\pi})$.
Subtracting these equalities
we obtain
$r_j-r_i=\nu(r_{j^\pi}-r_{i^\pi})$
and hence $\{i,j\}=\{i^\pi,j^\pi\}$.
As a consequence for the third element
$k\in \{1,2,3\}\setminus\{i,j\}$
we have $k^\pi=k$. If $p'\neq p$ 
then $p'-r_k=r_k-p$ (that is, $p'=2rk-p$)
and regarding $p'-r_i$ and $p-r_j$ we are left with the
following cases.
\\
(1) $\nu=-1$ and $\pi=1$ is the identity
\\
(2) $\nu=1$ and $\pi=(12)$
\\
(3) $\nu=-1$ and $\pi=(12)$.
\\
In case (3) we have 
$p-r_i=r_j-p'$ and $p-r_j=r_i-p'$,
that is, $p'=r_i+r_j-p$. Comparing this with
$p'=2r_k-p$ we obtain $2r_k=r_i+r_j$
and hence $r_i-r_k=r_k-r_j$, a contradiction
with our assumptions.
\\
In case (1) we obtain $p'=2r_i-p=2r_j-p=2r_k-p$,
which is solvable for every $p$ iff
$2r_i=2r_j=2r_k$.
\\
In case (2) we obtain $p'=p-r_i+r_j=p-r_j+r_i=p-2r_k$
which is solvable iff $2r_i=2r_j$ and $2p=2r_k-r_i+r_j$.
\end{proof}

\begin{proposition}
\label{line-prop}
Let $G$ be an abelian group
and assume that there is
an epimorphism $\phi:G\rightarrow \Z_4$.
We set $D=G$, $R=\{r_1,r_2,r_3\}$
and $P=\phi^{-1}(0)
\setminus \{p\in G| 2p=r_i+r_j\mbox{~for some~}i,j=1,2,3\}$,
where $\phi(R)=\{1,3\}$, and
$r_i-r_j\neq r_{i'}-r_{j'}$ whenever
$i\neq j$ and $(i,j)\neq (i',j')$.
Then
$(P,R)$ is a strictly addressable scheme
in $G$.
\end{proposition}

\begin{proof}{Proof}
Let $p\in P$, $p'\in G$,
$r_1',r_2',r_3'\in P\cup R$
such that $p'-r_i'=\pm (p-r_i)$.
Observe that $\{\phi(p-r_i)|i=1,2,3\}=\{1,3\}$.
Therefore $\{\phi(p'-r_i)|i=1,2,3\}=\{1,3\}$ as well.
As $\phi(\{P\cup \{r_1,r_2,r_3\})=\{0,1,3\}$,
this is only possible if $\phi(p')$ is even
and $\phi(r_i')$ are odd ($i=1,2,3$). As
$P\subseteq \phi^{-1}(0)$, this implies
$\{r_1',r_2',r_3'\}\subseteq \{r_1,r_2,r_3\}$
whence, by Lemma~\ref{refs-lemma},
$p'=p$ and $r_i'=r_i$ ($i=1,2,3$).
\end{proof}

Proposition~\ref{line-prop} gives that
$(P_1,R_1)$ is a strictly addressable scheme
in the group $\Z$, where $R_1=\{1,3,9\}$ and 
$P_1=4\Z$. In $\Z_4m$ for $m>2$ we obtain
the strictly addressable scheme
$(\Z_{4m}\setminus \{2m+2,2m+6\},\{1,3,9\})$.

We shall make use of the following
lemma in Section~\ref{shift-sect}.

\begin{lemma}
\label{line4-lemma}
Let $G$ be an abelian group
and assume that there is
an epimorphism $\phi:G\rightarrow \Z_4$.
Let $D=G$ and let $r_1,r_2,r_3,r_4\in \phi^{-1}\{1,3\}$
such that $\phi(\{r_1,r_2,r_3\})=\{1,3\}$,
$3r_4\neq r_1+r_2+r_3$, and
$r_i-r_j\neq r_{i'}-r_{j'}$ whenever
$i\neq j$ and $(i,j)\neq (i',j')$.
Then with
$P=\phi^{-1}(0)$, the element $r_4$ is
strictly addressable by $\{r_1,r_2,r_3\}$
with respect to the scheme
$(P\cup\{r_4\},\{r_1,r_2,r_3\})$.
\end{lemma}

\begin{proof}{Proof}
Assume that for some $r_4'\in G$,
for $r_1',r_2',r_3'\in P\cup\{r_1,r_2,r_3,r_4\}$
we have $r_4'-r_i'=\pm(r_4-r_i)$ ($i=1,2,3$).
We have $\{\phi(r_4-r_i)|i=1,2,3\}=\{0,2\}$,
whence also
$\{\phi(r_4'-r_i')|i=1,2,3\}=\{0,2\}$.
Since $\phi(P)=\{0\}$ and $\phi\{r_1,r_2,r_3,r_4\}=\{1,3\}$
this implies that $\phi(r_4')$ is odd and
$r_1',r_2',r_3'\in \{r_1,r_2,r_3\}$.
There exist $\mu=\pm 1$ and $i\neq j\in\{1,2,3\}$
such that $r_4'-r_i'=\mu(r_4-r_i)$ and
$r_4'-r_j'=\mu(r_4-r_j)$. As a consequence,
$r_i'-r_j'=\mu(r_i-r_j)$, whence either
$\mu=1$, $r_i'=r_i$, $r_j'=r_j$
or
$\mu=-1$, $r_i'=r_j$, $r_j'=r_i$.
In the first case we have $r_4'=r_4$
and, by the properties of $\{r_1,r_2,r_3,r_4\}$,
also $r_k'=r_k$ for $k=1,2,3$. Assume the second case.
Then $r_4'=r_i+r_j-r_4$. Let $k=\{1,2,3\}\setminus\{i,j\}$.
Then $r_i+r_j-r_4-r_k'=r_4'-r_k'=\nu(r_4-r_k)$ for some $\nu=\pm 1$.
If $\nu=-1$ then we obtain $r_i-r_k'=r_k-r_j$, a contradiction
to the properties of $\{r_1,r_2,r_3,r_4\}$. If
$\nu=1$ then $2r_4+r_k'=r_i+r_j+r_k$.
If $r_k'=r_k$, then we obtain $r_4-r_i=r_i-r_4$,
while if $r_k'=r_4$ then $3r_4=r_i+r_j+r_k=r_1+r_2+r_3$.
Both possibilities contradict to the properties
of $\{r_1,r_2,r_3,r_4\}$.
\end{proof}

\subsection{Schemes for Euclidean global gates in $\Z^s$}

In this subsection we show that strictly addressable
schemes $(P,R)$ exist in the lattice $\Z^s$ where
$|R|=\min\{3,s+1\}$ and $P$ has density $1/8^s$ 
in $\Z^s$. Here $\Omega=\R^s$, $D=\Z^s$,
$G=E(s)$ and the orbits of $G$ on pairs
correspond to distances. This is of the type
of our main geometric example for global gates
given in Section~\ref{globgates-sect}. 
For $s=1$ see  Proposition~\ref{line-prop}
gives a scheme of density $\frac{1}{4}$.
The optimality of the density $(1/8)^s$ (for $s>1$)
is an open question. Also note that the 
addressable schemes for the group $E(s)$ 
are also addressable schemes for the
subgroups of $E(s)$, most notably for the
the group of isometries preserving the
lattice $\Z^s$.

Assume that $s>1$. We shall actually show
that for an appropriate choice of an $s+1$-element
subset $R$ of $\Z^s$, and a set $P$ consisting of 
almost all points of the set $8\Z^s=\{(8z_1,\ldots,8z_s)|z_i\in \Z\}$,
the scheme $(P,R)$ is strictly addressable.
Let $k=s+2$ and for $i=1,\ldots, k$ 
write $P_i$ for the vector 
of variables $P_i=(x_{i1},x_{i2}, \ldots , x_{is})$.

We set $k=s+2$. A $k$-element 
subset $S\subset \R^s$ is {\em reconstructible}, if any other
$k$-element subset $S'\subset \R^s$ which has the same
distribution of pairwise distances among its points is actually isomorphic
to $S$ with respect to rigid motions  
(i.e., the group $E(s)$).

From a result of Boutin and Kemper (Theorem 2.6 in \cite{BK}) it follows 
that there
exists a nonzero polynomial $f(P_1, \ldots ,P_k)$ with real coefficients
such that if $p_1,p_2,\ldots ,p_k\in \R^s$ are points with
$f(p_1,\ldots, p_k)\not=0$ then $\{p_1,\ldots ,p_k \}$ is reconstructible.
The polynomial $f$ has the additional property (Lemma 2.9, loc. cit.) that
$f(p_1,\ldots, p_k)=0$ if the pairwise distances among the $p_i$ are not all
distinct.

If $0\not= g(x)\in \R[x]$ is a polynomial and $H$ is an infinite
subset of $\R$, then there exists a $c\in H$ such that $g(c)\not=0$.
A repeated application of this simple fact to the Boutin-Kemper polynomial 
shows the existence of a subset
$R=\{r_1,\ldots , r_{k-1}\} \subseteq  4\Z^s+(1,0,\dots,0)$ such that the
polynomial 
$$h(P_k):=f(r_1, r_2, \ldots ,r_{k-1}, P_k)$$ 
is not identically zero.
Moreover, we can suppose that  $r_1\in 8\Z^s+(1,0,\ldots , 0)$ and
$r_2\in 8\Z^s+(5,0,\ldots , 0)$. Let $P$ be the set of points $p$ from
$8\Z^s$  for which $h(p)\not=0$. Note that $P$ contains almost every
point of $8\Z^s$. Also, if $p\in P$ then the $s+1$ distances $|p-r_i|$ are 
all different, and the set of these distances uniquely determines $p$ among
the points of $\R^s$.

We now show that the above pair $(R,P)$ is a strictly addressable 
scheme on $D=\Z^s$.  We note first that for every $p\in P$
we have $|p-r_1|^2\equiv 1\pmod {16}$ while
$|p-r_2|^2\equiv 9\pmod {16} $. 
Assume that we have a $p\in P$, 
$q\in Z^s$ and $R'\subset  P\cup R$ 
such that 
$$ \{|p-r|, ~r\in R\}= \{|q-r'|, ~r'\in R'\}. $$
As for every $p_1,p_2\in P$ we have
$|q-p_1|^2\equiv |q-p_2|^2 \pmod {16}$, 
the fact
$|p-r_1|^2\not\equiv |p-r_2|^2\pmod {16}$
implies that $R'$ can not be 
entirely in $P$. Also, $q$ modulo 2 differs 
from $(0,\ldots,0)$ in an odd number of 
coordinates while it differs
from $(1,\ldots,0)$ in an even
number of coordinates or conversely.
Therefore $|q-p'|^2\not \equiv |q-r|^2\pmod 2$
if $p'\in P$ and $r\in R$. This
rules out the possibility that $R'$ intersects
properly both $P$ and $R$. It follows that
$R'\subseteq R$ and hence $R'=R$. 
Note that $R\cup\{p\}$ is rigid in the sense that
the pairwise distances among the points are all different
The re constructibility 
and rigidity of  $R\cup \{p\}$ gives that 
$q=p$ and hence the claim.

\section{A scheme with shiftable base}

\label{shift-sect}

In this section we show that 
on any finite interval of $\Z$ of
length $n$, a simulation of an
$\lfloor n/4 \rfloor$-qubit circuit
can be done efficiently using global one-qubit gates
and global two-qubit gates
with Hamiltonians of the form $A^{\orb_j}$
with $O<|j|\leq 22$.

Here $\Omega=\Z$ and $G=\Z$
acts on $\Omega$ by translation.
Then $\Orbits=\{\orb_j|j\in \Z\}\setminus\{\orb_0\}$
where $\orb_j=\{(u,v)\in D\times D\mid u-v=j\}$.
Let $P=4\Z$, 
and for every $\ell\in \Z$
let $R_\ell=\{4\ell+1,4\ell+3,4\ell+9$\}.
Just like in the argument preceding Corollary~\ref{main-coro}, 
it is sufficient to show how to
simulate local one-qubit gates at point of $P$
because the interaction graph of the global 
2-qubit Hamiltonian $A^{\orb_4}$ connects
every pair of points in $P$. 

During simulation of a one-qubit operation
at position $4\ell$, the three qubits at 
positions from $R_\ell$
are set to one and the qubits at position
from $\Z\setminus (P\cup R_\ell)$ are set to zero.
This is a state admitted by the scheme $(P,R_\ell)$.
By Proposition~\ref{line-prop}, $(P,R_\ell)$
is a strictly addressable scheme. In particular,
$4\ell$ and $4\ell+4$ are $3$-addressable.
Therefore, by Theorem~\ref{simu-thm}, any one-qubit gate can be 
efficiently approximated using global one-qubit gates
and global two-qubit gates with
Hamiltonians of the form $A^{\orb_j}$
with $O<|j|\leq 9$.

After each step of simulation, we have
a state admitted by $(P,R_\ell)$ for some
$\ell\in\Z$. (Initially $\ell=0$.) To perform
another simulation step, i.e., to simulate
a one-qubit gate at position $4\ell'$,
we have to produce the state admitted by
$(P,R_{\ell'})$ where the state of the qubits
at positions from $P$ are left unchanged.
Intuitively, the configuration $R_\ell$
has to be shifted to $R_{\ell'}$. 
Below we show a procedure doing this for
$\ell'=\ell+1$. For general $\ell'$
we need $|\ell-\ell'|$ iterations of
this basic procedure or its inverse.

In every intermediate step during the 
shifting procedure the qubits at
three positions $\{r_1,r_2,r_3\}$
are set to one while the value of
the qubit at a fourth position $r_4$
is flipped. In each step the system
$\{r_1,r_2,r_3\},r_4$ satisfies the
conditions of Lemma~\ref{line4-lemma}
and also $|r_i-r_j|\leq 22$. Therefore
the one-qubit operation flipping
the value of the qubit at $r_4$
can be efficiently approximated using
global one-qubit gates and 
and global two-qubit gates
with Hamiltonians of the form $A^{\orb_j}$
with $O<|j|\leq 22$.
Below we give the sequence of consecutive
configurations $\{r_1,r_2,r_3\},r_4$.

\begin{enumerate}
\item[Step 1.] $\{r_1,r_2,r_3\}=\{4\ell+1,4\ell+3,4\ell+9\}$, $r_4=4\ell+23$ ($r_4:0\mapsto 1$).
\\ {\scriptsize (Here $\{|r_i-r_j|\}=\{2,6,8\}\cup\{22,20,14\}$,
$r_1+r_2+r_3=12\ell+16\neq 12\ell+69=3r_4$.)}
\item[Step 2.] $\{r_1,r_2,r_3\}=\{4\ell+3,4\ell+9,4\ell+23\}$, $r_4=4\ell+1$ ($r_4:1\mapsto 0$).
\\ {\scriptsize (Here $\{|r_i-r_j|\}=\{6,14,20\}\cup\{2,8,22\}$,
$r_1+r_2+r_3=12\ell+35\neq 12\ell+3=3r_4$.)}
\item[Step 3.] $\{r_1,r_2,r_3\}=\{4\ell+3,4\ell+9,4\ell+23\}$, $r_4=4\ell+5$ ($r_4:0\mapsto 1$).
\\ {\scriptsize (Here $\{|r_i-r_j|\}=\{6,14,20\}\cup\{2,4,18\}$,
$r_1+r_2+r_3=12\ell+35\neq 12\ell+15=3r_4$.)}
\item[Step 4.] $\{r_1,r_2,r_3\}=\{4\ell+3,4\ell+5,4\ell+23\}$, $r_4=4\ell+9$ ($r_4:1\mapsto 0$).
\\ {\scriptsize (Here $\{|r_i-r_j|\}=\{2,18,20\}\cup\{6,4,14\}$,
$r_1+r_2+r_3=12\ell+31\neq 12\ell+27=3r_4$.)}
\item[Step 5.] $\{r_1,r_2,r_3\}=\{4\ell+3,4\ell+5,4\ell+23\}$, $r_4=4\ell+11$ ($r_4:0\mapsto 1$).
\\ {\scriptsize (Here $\{|r_i-r_j|\}=\{2,18,20\}\cup\{8,6,12\}$,
$r_1+r_2+r_3=12\ell+31\neq 12\ell+23=3r_4$.)}
\item[Step 6.] $\{r_1,r_2,r_3\}=\{4\ell+3,4\ell+5,4\ell+11\}$, $r_4=4\ell+23$ ($r_4:1\mapsto 0$).
\\ {\scriptsize (Here $\{|r_i-r_j|\}=\{2,6,8\}\cup\{20,18,12\}$,
$r_1+r_2+r_3=12\ell+16\neq 12\ell+69=3r_4$.)}
\end{enumerate}
Steps 7--12 are the same as Steps 1--6, respectively, with
$r_i+2$ in place of $r_i$.

\bigskip

{\bf Acknowledgements:} 
The authors are grateful to two anonymous referees 
for invalueble remarks and suggestions. We acknowledge 
financial support by project RESQ IST-2001-37559 of the 
IST-FET program of the EC, and 
by the Hungarian Scientific Research Fund (OTKA) under
grants T42706 and T42481.

\end{document}